\documentclass{iaus}
\usepackage{graphics,epsfig}

  \checkfont{eurm10}
  \iffontfound
    \IfFileExists{upmath.sty}
      {\typeout{^^JFound AMS Euler Roman fonts on the system,
                   using the 'upmath' package.^^J}%
       \usepackage{upmath}}
      {\typeout{^^JFound AMS Euler Roman fonts on the system, but you
                   dont seem to have the}%
       \typeout{'upmath' package installed. iaus.cls can take advantage
                 of these fonts,^^Jif you use 'upmath' package.^^J}%
      }
  \else
  \fi

  \checkfont{msam10}
  \iffontfound
    \IfFileExists{amssymb.sty}
      {\typeout{^^JFound AMS Symbol fonts on the system, using the
                'amssymb' package.^^J}%
       \usepackage{amssymb}%

      }{}
  \fi

  \IfFileExists{amsbsy.sty}
    {\typeout{^^JFound the 'amsbsy' package on the system, using it.^^J}%
     \usepackage{amsbsy}}
    {\providecommand\boldsymbol[1]{\mbox{\boldmath $##1$}}}

\newsavebox{\astrutbox}
\sbox{\astrutbox}{\rule[-5pt]{0pt}{20pt}}

\title[Outskirts of Galaxy Clusters: intense life in the suburbs]
      {A note on Temperature Profiles of rich Clusters of Galaxies}

\author[ De Grandi \& Molendi]%
{Sabrina De Grandi$^1$%
\and Silvano Molendi $^2$}

\affiliation{$^1$ INAF OAB, via E. Bianchi 46, 23807 Merate (LC),
Italy
email: degrandi@mi.astro.it\\[\affilskip]
$^2$ IASF, Sez. di Milano, Via Bassini 15, I-20133 Milano Italy
email: silvano@mi.iasf.cnr.it}

\pubyear{2004} \volume{195} \pagerange{}
\date{}
\jname{Outskirts of Galaxy Clusters: intense life in the suburbs}
\editors{A. Diaferio, ed.}
\begin{document}

\maketitle

\begin{abstract}
We  derive here the  mean temperature profile  for  a sample   of
hot, medium distant clusters recently observed with XMM-Newton,
whose  profiles are available from  the literature, and  compare
it with  the mean temperature profile found from BeppoSAX data.
The XMM-Newton and BeppoSAX profiles are in good agreement between
0.05 and 0.25 $r_{180}$. From 0.25 to about 0.5 $r_{180}$ both
profiles decline, however the BeppoSAX profile does so much more
rapidly than the XMM-Newton profile.\end{abstract}

\firstsection 
\section{Introduction}

Temperature profiles of  galaxy clusters are  of great  importance for
two main reasons: firstly they  allow us to measure  the total mass of
these  systems  through   the   hydrostatic  equilibrium equation  and
secondly they provide informations on  the thermodynamic state of  the
Intra Cluster  Medium  (hereafter ICM).

Measurements of temperature profiles have  been performed with the
the first generation of X-ray  satellites carrying telescopes
operating in the medium energy band (2-10 keV), namely  ASCA and
BeppoSAX. A detailed description of results obtained with these
experiments may be found in De Grandi \& Molendi (2002) and refs.
therein.

In  the  last   3 years   various  authors have  published
XMM-Newton temperature profiles  of individual clusters. Arnaud et
al. (2003) with a  sample  of 7 objects comprising 5 clusters and
2 groups  find that  the temperature profiles are essentially
isothermal within $0.5~r_{200}$, and possibly declining at larger
radii, where the statistics is rather limited.  Recently, Zhang
et al. (2003)  have published temperature profiles for  a sample
of 9 clusters, in the  outer  regions they find  both flat and
strongly  decreasing profiles, however they  do not provide  a
mean temperature profiles of their sample. In this note we derive
the mean temperature profile for  a sample of hot, medium distant
clusters whose profiles are  available from the literature and
compare it with our mean BeppoSAX profile.

\section{XMM-Newton sample selection from the literature}\label{sec:sample}

We have  selected  from the  literature   all hot (i.e.   kT$>3$
keV) clusters in the  redshift range between 0.1  and 0.3 with an
available projected radial temperature profile. The  resulting
sample  comprises a total    of 15 clusters: 9  REFLEX clusters at
redshift $\sim 0.3$ (Zhang et al.  2003), A1413 at z=0.143 (Pratt
\& Arnaud 2002), A2163  at z=0.201 (Pratt  et al.  2001), A1835 at
z=0.250 (Majerovitz et al.  2002), PKS 0745$-$191 at z=0.1028
(Chen et al.  2003),  ZW 3146  at z=0.291 and   E1455$+$223 at
z=0.258  both taken from Mushotzky  (2003). The adopted redshift
range allows us  to  compare the temperature profiles measured
from  XMM-Newton data with those derived    from BeppoSAX observations.
The profiles  published for  the  clusters in the XMM-Newton sample
extend out to $\sim 9^\prime-10^\prime$ from the core,
corresponding to physical radii of $\sim 1.5-3$ Mpc (H$_0 = 50$ km
s$^{-1}$ Mpc$^{-1}$).  The same physical radii are reached by
BeppoSAX observations of clusters laying in the $\sim 0.05-0.08$
redshift range and detected out to $\sim 20^\prime$ (see De Grandi
\& Molendi 2002). The average cluster temperature of  the
XMM-Newton sample is $\sim$ 7 keV, very similar to that of the
BeppoSAX sample ($\sim 6$ keV). We have converted all  temperature
uncertainties  at the  90\% confidence level into errors at the
68\% c.l.  by dividing them by the scaling  factor 1.65.  For each
cluster we have  computed the virial radius $r_{vir}$ ($=
r_{180}$) from  the relation derived by Evrard  et al.   (1996):
$r_{vir}   =  3.9   ~ \sqrt{T\over   {10~{\rm  keV}}}
~(1+z)^{-3/2}~~~{\rm Mpc},$   using published cluster mean
temperatures and redshifts. 

\begin{figure}
\centering
  \includegraphics[angle=-90,width=1.0\hsize]{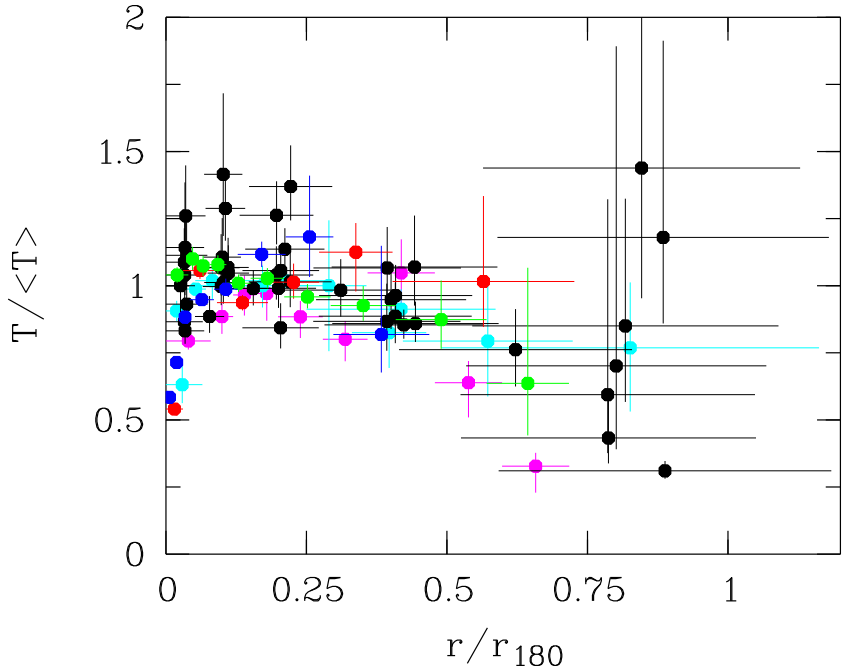}
  \caption{Projected temperature profile for 15 clusters
observed with EPIC, radius in units of r$_180$. Profiles are from
the literature (references in text).
Errors are at the $68\%$ c.l.} \label{fig:profiles}
\end{figure}

Visual inspection of Fig.1, where we report the individual XMM-Newton
temperature profiles, shows that the profiles are about isothermal
from 0.10 to $0.3-0.5~r_{180}$; beyond $0.3-0.5~r_{180}$ there
seems to be a decline. It is also clear that points at
$r_{180}>0.6-0.7$ are heavily fluctuating and tend to have larger
errors.

We have modelled the XMM-Newton profiles in Fig. 1 with a power law and
converted the slope into the polytrophic gas index accordingly to
the calculation reported in the appendix of De Grandi \& Molendi
(2002). In the range $0.2 < r_{180} < 0.5$ we have obtained a
polytrophic index $\gamma = 1.09\pm0.03$, which is close to the
isothermal value 1. Whereas, for $0.5 < $r$_{180} < 0.75$ we have
found $\gamma = 2.4 \pm 0.5$, which is formally above the
adiabatic limit of 5/3, although consistent at the 68\% confidence
level with values below 5/3. The drop in the profiles at $\sim$
0.55 r$_{180}$ (clearly visible in Fig. 2) highlights  a possible
problem in the XMM-Newton profiles.

\section{The mean profile}

The binned  error-weighted average  temperature profile computed
from the 15 clusters is  shown in Fig. 2, where we also plot the
BeppoSAX cool core and non cool core clusters profiles (see De
Grandi \& Molendi 2002 for details). We have added here to the
BeppoSAX cool core clusters the Ophiuchus cluster.

\begin{figure}
\centering
 \includegraphics[angle=-90,width=1.0\hsize]{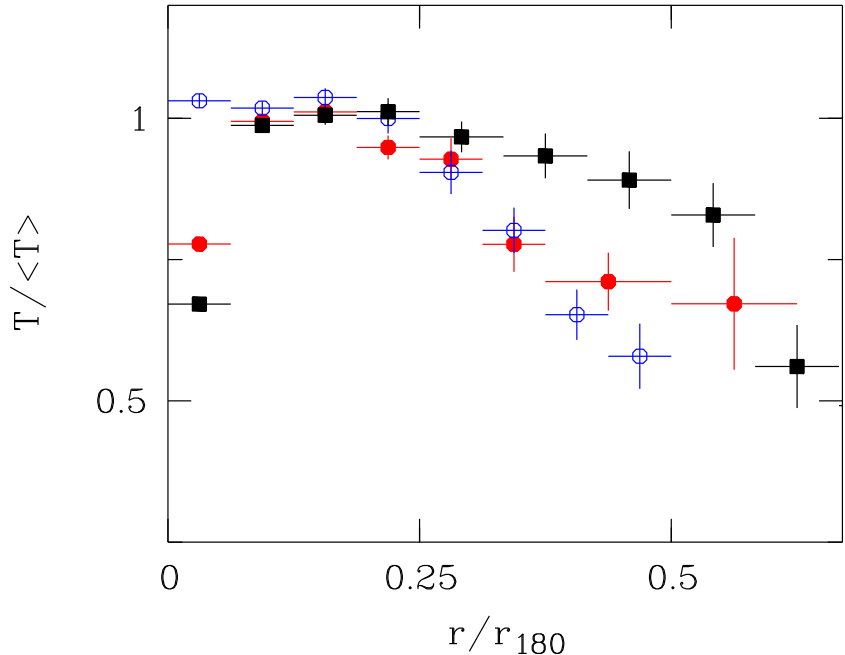}
  \caption{Mean temperature profile from BeppoSAX (De Grandi \& Molendi 2002)
(full and empty circles are respectively cool core and non cool
core clusters) and from XMM-Newton (squares).}\label{fig:mean}
\end{figure}

The XMM-Newton and BeppoSAX profiles are in good agreement between 0.05
and 0.25 $r_{180}$. From 0.25 to about 0.5 $r_{180}$ both profiles
decline, however the BeppoSAX profile does so more rapidly than
the XMM-Newton profile. The presence of a temperature jump, at the
limit of  convective stability, at  the largest radii explored
with XMM-Newton, hints to a problem possibly related to the low
signal-to-noise ratio of the outermost cluster regions and to the
difficulties in operating a correct background subtraction (see
Molendi these proceedings for a discussion of this issue).



\begin{thebibliography}{}

   \bibitem[]{} {Arnaud, M., Pratt, G. W. \& Pointecouteau, E.} 2004,
   Mem. S.A.It. astro-ph/0312398.

   \bibitem[]{} {Chen, Y., Ikebe, Y. \& Boehringer, H.} 2003, A\&A, \textbf{407}, 41.

   \bibitem[]{} {De Grandi, S. \& Molendi, S.} 2002, ApJ, \textbf{567}, 163.


   \bibitem[]{} {Majerowicz, S., Neumann, D. M. \& Reiprich, T. H.} 2002,
   A\&A, \textbf{394}, 77.


   \bibitem[]{} {Molendi, S.} 2004, these proceedings.

   \bibitem[]{} {Mushotzky, R. F.} 2003, astro-ph/0311105.

   \bibitem[]{} {Pratt, G. W. \& Arnaud, M.} 2002, A\&A, \textbf{394}, 375.


   \bibitem[]{} {Pratt, G. W. et al.} 2001, Proc. of  
   XXIth Moriond Astroph. Meeting, p.38.


   \bibitem[]{} {Zhang, Y.-Y., Finoguenov, A., Boehringer, H., et al.} 
   2004, A\&A, \textbf{413}, 49.

\end{thebibliography}
\end{document}